# Crystal structure and low-energy Einstein mode in $ErV_2Al_{20}$ intermetallic cage compound

Michał J. Winiarski[1,*], Tomasz Klimczuk[1]

[1] *Faculty of Applied Physics and Mathematics, Gdansk University of Technology, Narutowicza 11/12, 80-233 Gdansk, Poland*

\* *corresponding author: mwiniarski@mif.pg.gda.pl*

**Abstract**

Single crystals of a new ternary aluminide $ErV_2Al_{20}$ were grown using a self-flux method. The crystal structure was determined by powder X-ray diffraction measurements and Rietveld refinement, and physical properties were studied by means of electrical resistivity, magnetic susceptibility and specific heat measurements. These measurements reveal that $ErV_2Al_{20}$ is a Curie-Weiss paramagnet down to 1.95 K with an effective magnetic moment $\mu_{eff}$ = 9.27(1) $\mu_B$ and Curie-Weiss temperature $\Theta_{CW}$ = -0.55(4) K. The heat capacity measurements show a broad anomaly at low temperatures that is attributed to the presence of a low-energy Einstein mode with characteristic temperature $\Theta_E$ = 44 K, approximately twice as high as in the isostructural 'Einstein solid' $VAl_{10.1}$.

**Introduction**

Ternary aluminides $RT_2Al_{20}$ (R = electropositive elements, T – early 3*d*, 4*d*, and 5*d* transition metals) crystallizing in the $CeCr_2Al_{20}$-type structure have recently attracted much scientific interest, due to their versatility towards the chemical composition and a variety of interesting physical properties, depending on the constituent elements.

The unit cell of a $CeCr_2Al_{20}$-type compound is shown in Figure 1. A characteristic feature of this structure is the presence of large 'cages' formed by Al atoms surrounding the electropositive *R* atom (occupying the Wyckoff position 8*a*). Such cages of various shapes and sizes are also found eg. in dodecaboride $ZrB_{12}$ [1], clathrates including $Ba_8Si_{46}$, $Ba_{24}Si_{100}$ [2], $Ba_8Au_{16}P_{30}$ [3], $Na_{24}Si_{136}$ [4], filled skutterudites [5–7] and in the β-pyrochlore oxides $AOs_2O_6$ (*A* = alkaline metals) [8] (see Figure 2). In these systems low-energy, large amplitude localized (related to a specific crystallographic site) anharmonic vibrations of small cage-filling atoms are observed. This 'rattling' of cage filling atoms affects thermal and transport properties of materials, resulting eg. in suppression of the lattice thermal conductivity [3,5,9] that can lead to an increase in the thermoelectric figure of merit (ZT) [3,5]. The presence of 'rattling' phonons is also found to enhance the superconducting critical temperature in

hexa- and dodecaborides [1,10,11], β-pyrochlores [8,12], and CeCr$_2$Al$_{20}$-type compounds: Al$_x$V$_2$Al$_{20}$ [13–15], Ga$_x$V$_2$Al$_{20}$ [14,16], and $R$V$_2$Al$_{20}$ ($R$ = Sc, Y, Lu) [17]. The contribution of 'rattling' phonons to the specific heat are usually well described using the Einstein model with a low characteristic (Einstein) temperature $\Theta_E$.

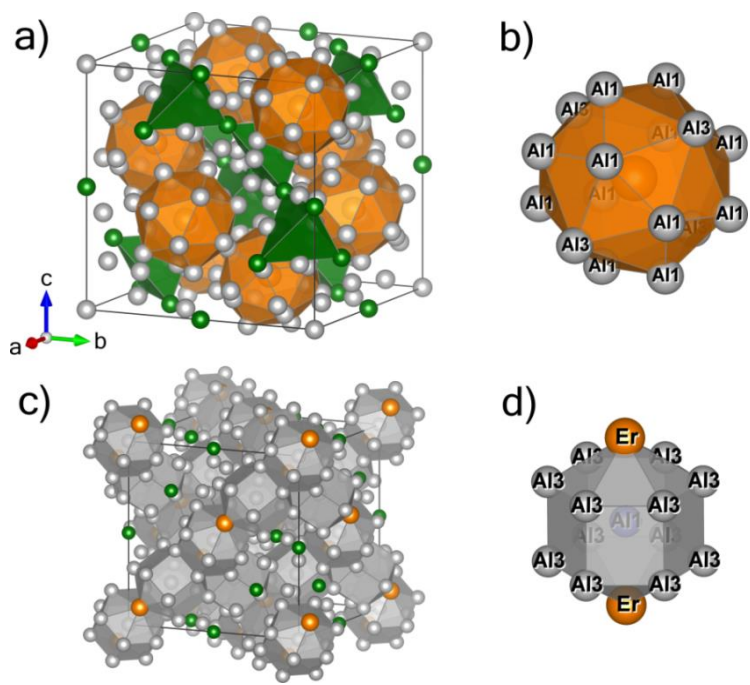

**Figure 1 Unit cell of ErV$_2$Al$_{20}$.** (a) Er atoms (orange), arranged in a diamond lattice, are positioned inside Frank-Kasper CN 16 polyhedra [18] (Panel b) formed by Al1(16$c$) and Al3(96$g$) atoms (silver). V atoms (green) form a pyrochlore array [19]. (c,d) Al1(16$c$) atoms are encaged in Frank-Kasper CN 14 polyhedra formed by Al3(96$g$) and Er(8$a$). Image was rendered using VESTA software [20]. Fig. S1 of Supplementary Material shows the relationships between the CeCr$_2$Al$_{20}$, ZrZn$_{22}$, and Mg$_3$Cr$_2$Al$_{20}$-type structures.

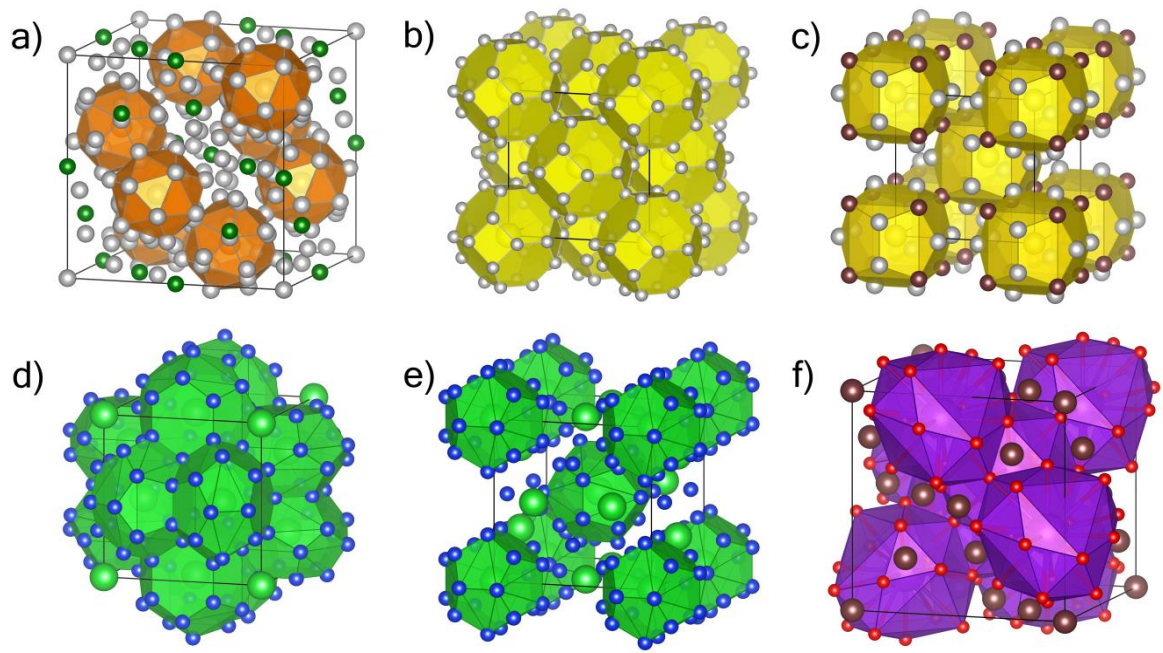

**Figure 2 Comparison of cage-type crystal structures:** (a) CeCr$_2$Al$_{20}$-type ErV$_2$Al$_{20}$, (b) ZrB$_{12}$ dodecaboride (Zr atoms – yellow, B – silver) [21,22], (c) PrOs$_4$Sb$_{12}$ skutterudite (Pr – yellow, Os – brown, Sb – silver) [23,24], (d,e) Ba$_8$Si$_{46}$ clathrate (Ba –green, Si – blue) [22,25], with two distinct Ba-Si cage types, (f) KOs$_2$O$_6$ β-pyrochlore (K – purple, Os – brown, O – red) [22,26].

The introduction of 4$f$ elements into the 8$a$ position of a CeCr$_2$Al$_{20}$ structure leads to a variety of magnetic properties. LaV$_2$Al$_{20}$ was found to exhibit strong diamagnetism [27] resulting from a peculiar Fermi surface [28]. Ce$T_2$Al$_{20}$ ($T$ = Ti, V, Cr) are found to exhibit Pauli paramagnetism, consistent with a nonmagnetic tetravalent configuration of the Ce atom [29,30]. Pr$T_2$Al$_{20}$ ($T$ = Ti, V) are exotic superconductors in which the superconducting state coexists with quadrupolar order [13–16], while PrCr$_2$Al$_{20}$ was described as a paramagnetic Kondo lattice [30]. NdTi$_2$Al$_{20}$ is an antiferromagnet (AFM) with the Néel temperature $T_N$ = 1.45 K [35], NdV$_2$Al$_{20}$ and NdCr$_2$Al$_{20}$ in turn are ferromagnets (FM) with the Curie temperature $T_C$ = 1.8 K and 1.7 K, respectively [35]. In Sm$T_2$Al$_{20}$ ($T$ = Ti, V, and Cr) AFM transition ($T_N$ = 6.4, 2.9 and 1.8 K, respectively) was observed along with strong valence fluctuations and a Kondo effect [36], similarly SmTa$_2$Al$_{20}$ shows strong electron correlation effects [37]. EuV$_2$Al$_{20}$ was found to exhibit a Kondo lattice behavior [38] and EuCr$_2$Al$_{20}$ shows an AFM transition ($T_N$ = 4.8 K) due to the magnetic moments carried on in a divalent Eu ion [39]. GdTi$_2$Al$_{20}$, GdV$_2$Al$_{20}$, and GdCr$_2$Al$_{20}$ are reported AFMs with $T_N$ = 2.6, 2.4-3.1, and 3.9 K [40,41], respectively. TmTi$_2$Al$_{20}$ was recently found to exhibit an AFM transition at 0.7 K [42], while no magnetic ordering is observed in TmV$_2$Al$_{20}$ down to 0.5 K [43]. Yb$T_2$Al$_{20}$ ($T$ = Ti, V, Cr) are Pauli paramagnets with a divalent Yb ion [44] and LuV$_2$Al$_{20}$ was found to exhibit a SC transition with the critical temperature $T_c$ = 0.6 K [17]. Although some of the Tb, Dy, Ho, and Er-bearing compounds are reported [45], their physical properties remain unknown.

In this study we present single-crystal growth and physical characterization of a previously unreported ErV$_2$Al$_{20}$ intermetallic. The crystal structure of the new compound is described along with results of magnetic susceptibility, electrical resistivity and specific heat measurements.

**Materials and Methods**

Single crystals of ErV$_2$Al$_{20}$ were grown using an Al self-flux method [46]. Erbium (99.9% purity), vanadium (99.8%), and aluminum (99.99%) pure metals were put together in an alumina crucible at the atomic ratio of 1:2:90 (Er:V:Al). A frit-disc and a second crucible were used for flux separation as it is described in ref. [47]. The crucible set was then sealed in an evacuated quartz tube backfilled with Ar to dilute the Al vapor attacking the tube walls.

The ampoule was then placed in a furnace, heated at 100°C/h to 1070°C, held for 2 h, and then slowly cooled (4°C/h) to 770°C at which temperature it was centrifuged to separate crystals from the remaining flux. Crystals with sizes up to a few millimeters were obtained.

Crystals were then etched in ca. 0.1 M sodium hydroxide (NaOH) solution for a few hours to remove the Al flux droplets that remained after centrifugation.

Powder X-ray diffraction (XRD) patterns were collected on pulverized single-crystals using PANalytical X'Pert Pro diffractometer with Cu K$_\alpha$ source. To prevent the effect of preferred orientation along the [111] direction several small single crystals were first fine ground using an agate mortar and pestle and the resulting powder was spilled onto a spot of Apiezon M grease on a sample holder. FullProf software package [48] was used for the Rietveld refinement [49] of the structure model derived from crystallographic data for GdV$_2$Al$_{20}$ [41].

Resistivity measurements were carried out by the four contact method in a Quantum Design Physical Properties Measurement System (PPMS) on a single crystal sample cut and polished into a rectangular plate. Electrical leads were made of Ø0.04 mm platinum wires glued to the sample surface using a silver paste. Magnetic susceptibility measurements were done on etched single crystals using the ACMS option of PPMS in a temperature range of 1.95-300 K. Several randomly-oriented single crystals were taken and put in standard straw sample holders. Specific heat measurements were carried out using the PPMS Heat Capacity option by means of the standard 2τ relaxation method.

**Results**

*Crystallographic structure*

Figure 3 presents the powder XRD pattern collected on pulverized single crystals of ErV$_2$Al$_{20}$ along with a Rietveld fit to the data. Analysis of the diffraction reflections showed that the sample contained approx. 2 wt.% of Al that remained after etching in NaOH solution.

Table 1 presents the crystallographic structure parameters derived from the Rietveld fit. Lattice constant of ErV$_2$Al$_{20}$ $a$ = 14.5175(2) Å was found to be slightly larger than reported for TmV$_2$Al$_{20}$ (14.5024 Å) [43] and LuV$_2$Al$_{20}$ (14.5130 Å) and smaller than for DyV$_2$Al$_{20}$ (14.54 Å) [45], in agreement with the lanthanide contraction effect (see Figure 4). The unit cell parameter of ErV$_2$Al$_{20}$ was found to be larger than in ErCr$_2$Al$_{20}$ and smaller than for ErTi$_2$Al$_{20}$ (see inset of Figure 4) in consistency with behavior of the whole CeCr$_2$Al$_{20}$-type family.

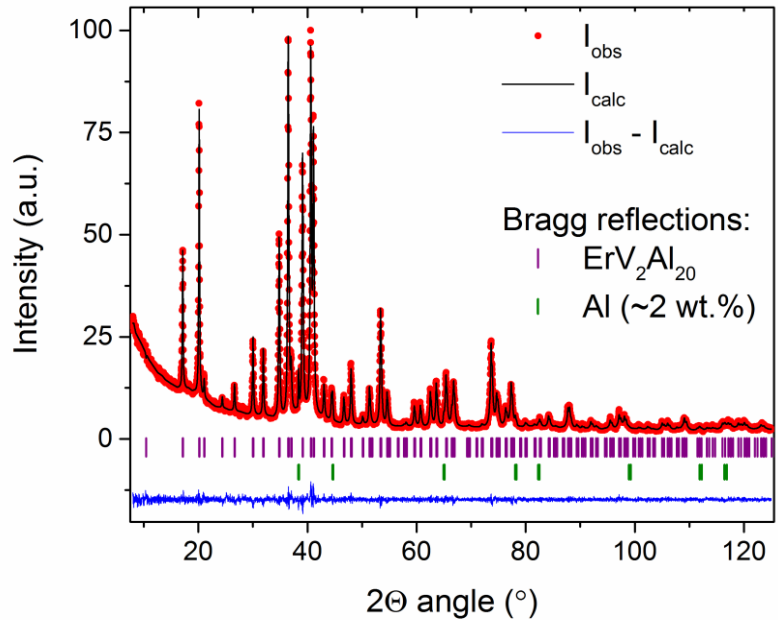

**Figure 3** Fit (black) to the experimental powder X-ray diffraction pattern of pulverized $ErV_2Al_{20}$ crystals (red points). The difference profile ($I_{obs} - I_{calc}$) is shown as a blue line. Purple and green ticks mark the positions of Bragg reflections for $ErV_2Al_{20}$ and Al impurity, respectively. The slightly higher background around $2\Theta \approx 40°$ comes from the Apiezon M grease used for sticking powder to the sample holder.

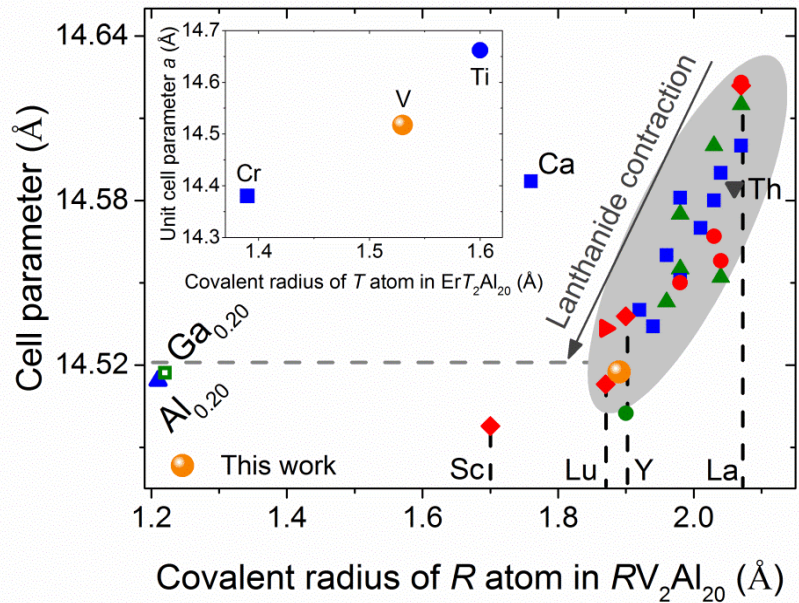

**Figure 4** The relation between unit cell parameter *a* and covalent radius of the cage-filling $R(8a)$ atom for the reported $RV_2Al_{20}$ compounds. Unit cell parameter decreases with increasing atomic number from La to Lu due to the lanthanide contraction effect. For atoms smaller than Lu, cell parameters does not change significantly, as it is discussed in ref. [17]. Inset shows unit cell parameters of $ErT_2Al_{20}$ ($T$ = Ti, V, Cr). Covalent radii of the elements are given after ref. [50], unit cell parameters: green triangles and blue squares – ref. [45] and references therein, red circles – ref. [30], red diamonds – ref. [17], red triangle ($YbV_2Al_{20}$) – ref. [44], blue triangle ($VAl_{10.1}$) – ref. [15], green circle ($TmV_2Al_{20}$) – ref. [43], green open square – ref. [14], grey triangle ($ThV_2Al_{20}$) – ref. [51], blue circle ($ErTi_2Al_{20}$) - [52].

**Table 1: Crystal structure parameters obtained from Rietveld fits to the XRD data collected at room temperature (ca. 20°C). Numbers in parentheses indicate statistical uncertainties of the least significant digits. The site occupancies were not relaxed during the refinement. The *R*-factors presented in the table are corrected for background contribution.**

|  |  | $ErV_2Al_{20}$ |
|---|---|---|
| Space group | | $Fd\bar{3}m$ (no. 227) |
| Z (number of formula units per unit cell) | | 8 |
| Pearson symbol | | *cF*184 |
| Cell parameter (Å) | | 14.5175(2) |
| Cell volume (Å$^3$) | | 3059.68(13) |
| Molar weight (g·mol$^{-1}$) | | 808.77 |
| Density (g·cm$^{-3}$) | | 3.51 |
| Er (8*a*) | x = y = z = | 1/8 |
| | $B_{iso}$ (Å$^2$) | 2.34(3) |
| V (16*d*) | x = y = z = | ½ |
| | $B_{iso}$ (Å$^2$) | 1.27(4) |
| Al1 (16*c*) | x = y = z = | 0 |
| | $B_{iso}$ (Å$^2$) | 2.42(10) |
| Al2 (48*f*) | x = | 0.4871(1) |
| | y = z = | 1/8 |
| | $B_{iso}$ (Å$^2$) | 1.61(6) |
| Al3 (96*g*) | x = y = | 0.0596(1) |
| | z = | 0.3240(1) |
| | $B_{iso}$ (Å$^2$) | 1.66(4) |
| Er-Er distance (Å) | | 6.2862(1) |
| Er-Al distances (Å): | Al1 | 3.143 |
| | Al3 | 3.186(2) |
| Figures of merit: | | |
| | $R_p$ (%) | 11.7 |
| | $R_{wp}$ (%) | 11.6 |
| | $R_{exp}$ (%) | 10.8 |
| | $\chi^2$ (%) | 1.16 |

*Electrical resistivity*

Resistivity of ErV$_2$Al$_{20}$ shows metallic-like character as it is shown in Figure 5(a). Since the single crystal used for measurements was cut and its surface was polished, the contribution of Al flux spots to the sample resistivity can be considered negligible. The low-temperature resistivity was found to show a complex character that cannot be fitted by simple models (see Figure 5(b)). The unusual T-dependence of resistivity results from the presence of a low-energy Einstein phonon mode (see the *Specific heat* paragraph below). It was shown by J.R. Cooper [53] that the contribution of an Einstein mode to the resistivity can be described using an equation:

$$\rho_{Einstein}(T) = \frac{KN}{MT(\exp\left(\frac{\Theta_E^*}{T}\right)-1)(1-\exp(\frac{-\Theta_E^*}{T}))} \quad (1)$$

where M is the mass of the oscillator, N is the number of oscillators per unit volume, K is a parameter dependent on the electron density and the strength of the coupling between electrons and local mode phonons, and $\Theta_E^*$ is the characteristic temperature of the Einstein mode (Einstein temperature). While this equation was successfully used for fitting the resistivity data of 'Einstein solids' VAl$_{10+x}$ and Ga$_x$V$_2$Al$_{20}$ [16,53], in the case of ErV$_2$Al$_{20}$ it was necessary to include additional terms, as is shown in Eq. 2:

$$\rho(T) = \rho_0 + \rho_{Einstein}(T) + AT^2 + BT^5 \quad (2)$$

where $\rho_0$ is a residual resistivity arising from both crystal and spin lattice disorder, BT$^5$ describes the resistivity resulting from electron-phonon scattering in the low-temperature limit and AT$^2$ accounts for electron-electron scattering. The fit yields the residual resistivity $\rho_0$ = 13 μΩ·cm, the electron-electron and electron-phonon scattering coefficients A = 2.7(4) · 10$^{-9}$ μΩ·cm K$^{-2}$ and B = 3.0(1) · 10$^{-9}$ μΩ·cm K$^{-5}$, and the parameters of scattering by Einstein mode phonons: $\Theta_E^*$ = 35(1) K and $\frac{KN}{M}$ = 97(5) μΩ·cm K (numbers in parentheses indicate the statistical uncertainties of the least significant digits).

The residual resistivity is of the same order of magnitude as reported for polycrystalline samples of YV$_2$Al$_{20}$, LaV$_2$Al$_{20}$ and LuV$_2$Al$_{20}$, and almost 10 times lower than for ScV$_2$Al$_{20}$, where the presence of structural disorder was speculated [17], however, when comparing $\rho_0$ of magnetic and non-magnetic compounds it is important to take into account the effect of spin disorder in the former. Taking the $\rho_0$ value from the fit, a residual resistivity ratio *RRR* = ρ(300 K)/$\rho_0$ ≈ 6.3 is calculated, suggesting a relatively good sample quality.

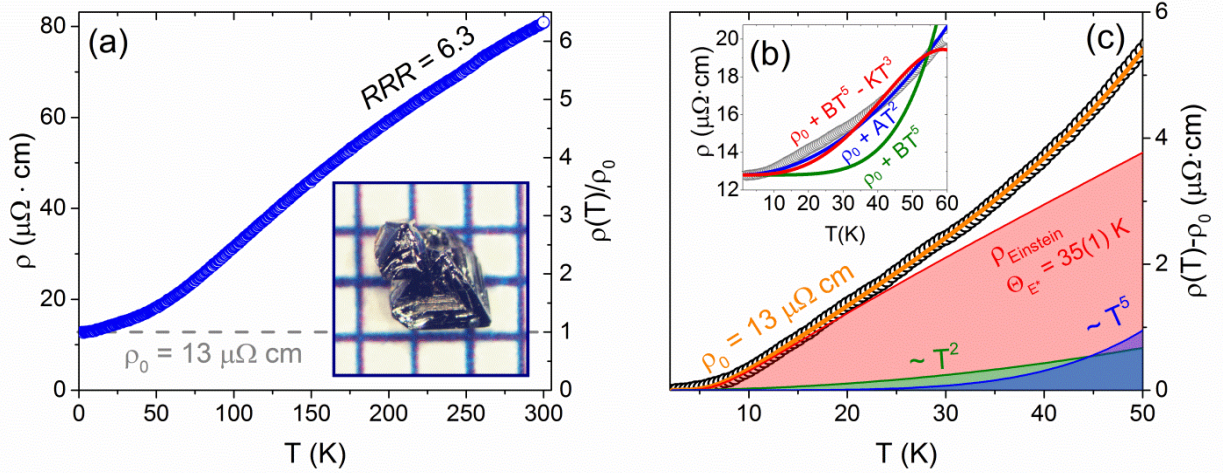

**Figure 5 (a):** Temperature dependency of resistivity of ErV$_2$Al$_{20}$. Residual resistivity ratio *RRR* = 6.3. Inset shows a sample aggregate of single crystals on a millimeter scale. **(b):** Low-temperature resistivity. Neither ~T$^2$ dependence (blue line), ~T$^5$ dependence (green line), nor ~T$^5$ combined with Mott ~T$^3$ term (red line) is sufficient to reproduce the experimental trend. **(c):** Fit (orange line) to the low-temperature resistivity (black circles) using Eq. 1, yielding residual resistivity $\rho_0$ = 13 μΩ·cm and the characteristic temperature of the Einstein mode $\Theta_E$ = 35(1) K. Red line show a contribution of Einstein mode to the resistivity, green of electron-electron scattering and blue of electron-phonon scattering.

*Magnetic susceptibility*

The temperature dependence of *dc* susceptibility χ(T) at 1 T field (Figure 6) is found to follow the Curie-Weiss law (eq. 3):

$$\chi(T) = \frac{C}{T - \theta_{C-W}} + \chi_0 \qquad (3)$$

where C is the Curie constant, $\theta_{C-W}$ is the Curie-Weiss temperature and $\chi_0$ is the temperature-independent contribution to the susceptibility (coming both from the sample and a sample holder). Figure 5(b) shows an inverse susceptibility vs. temperature plot emphasizing the Curie-Weiss character of magnetic susceptibility. The χ(T) data were fitted using the relationship given in Eq. 3. The Curie constant is related to the effective moment $\mu_{eff}$ associated with the magnetic ion as shown in eq. 4:

$$C = \frac{N_A \mu_B^2 \mu_{eff}^2}{3 k_B} \qquad (4)$$

where $N_A$ is the Avogadro number, $\mu_B$ – Bohr magneton, and $k_B$ – Boltzmann constant. Values extracted from the fit are gathered in Table 2. The obtained effective magnetic moment (9.27 $\mu_B$) is close to the value expected for a free Er$^{3+}$ ion (9.5 $\mu_B$) [54] and a small difference may be explained by both trace amounts of Al in the sample and effects of the crystal electric field. The close to zero yet negative value of $\Theta_{C-W}$ suggests the presence of only very weak effective interactions between Er$^{3+}$ magnetic moments. The field dependence of magnetization (see Fig. S2 of Supplementary Material) shows a saturating character at low temperature (2 K), as expected for a Curie-Weiss paramagnet. Results of *ac* magnetic susceptibility at low constant field $H_{dc}$ = 5 Oe ($B_{dc}$ = 0.5 mT) show no sign of a magnetic transition down to 1.95 K (see Figure 6(c)).

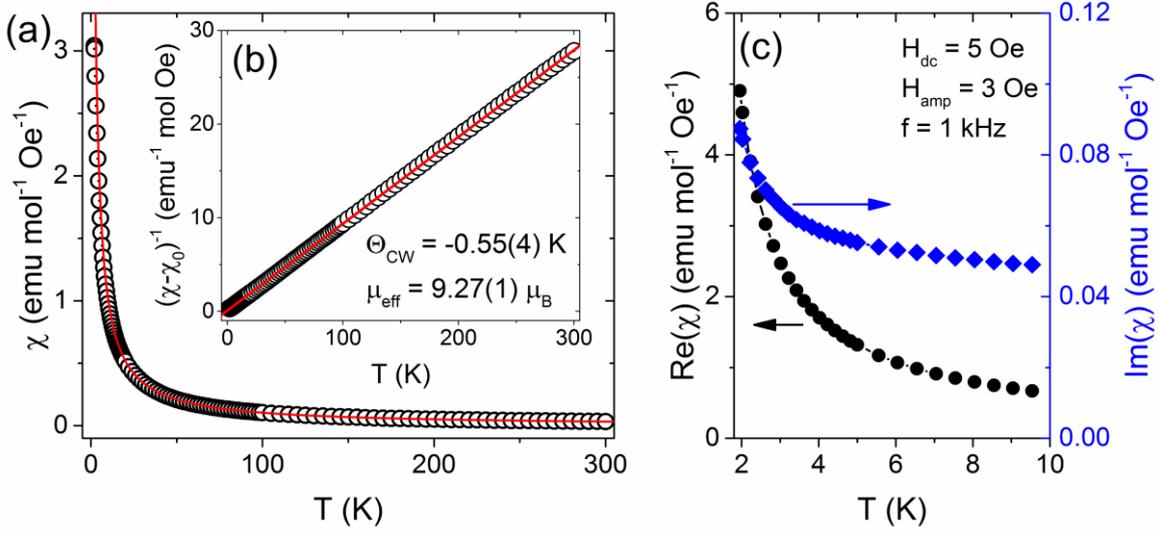

Figure 6 (a) Magnetic susceptibility of ErV$_2$Al$_{20}$. The Curie-Weiss fit (Eq. 3) to the experimental points (black circles) is shown as a red solid line. (b) Inverse susceptibility (corrected for the temperature-independent contribution) shows a linear relation consistent with the Curie-Weiss law down to the lowest temperatures measured (1.95 K). (c) *ac* magnetic susceptibility at low *dc* field H$_{dc}$ = 5 Oe. No magnetic transition is observed down to 1.95 K. The plot of magnetization vs. applied magnetic field at 2 K and 50 K is shown in Fig. S2 of the Supplementary Material.

Table 2: Results of Curie-Weiss fit to the *dc* susceptibility at 1 T. Note that the temperature-independent susceptibility ($\chi_0$) is not corrected for the contribution of sample holder.

| | |
|---|---|
| C (emu K mol$^{-1}$ Oe$^{-1}$) | 10.73(2) |
| $\mu_{eff}$ ($\mu_B$) | 9.27(1) |
| $\Theta_{CW}$ (K) | -0.55(4) |
| $\chi_0$ (emu mol$^{-1}$ Oe$^{-1}$) | -1.4(1) · 10$^3$ |

*Specific heat*

The specific heat of ErV$_2$Al$_{20}$ single crystal in the temperature range 1.95-30 K at zero magnetic field is shown in Figure 7. Two anomalies are seen in the C$_p$/T vs. T plot: the first one below 5 K, attributed to the Schottky anomaly and second between 5 and 20 K, attributed to a low-energy Einstein phonon mode, observed in the isostructural 'Einstein solid' VAl$_{10.1}$ [13,15].

The experimental data are fit with a function including electronic, phonon and Schottky contributions:

$$C_p = C_{electronic} + C_{Debye} + C_{Einstein} + C_{Schottky}$$

$$C_{electronic} = \gamma T$$

$$C_{Debye} = \beta T^3 \qquad (5)$$

$$C_{Einstein} = A \cdot 3nR \left(\frac{\Theta_E}{T}\right)^2 \exp\left(\frac{\Theta_E}{T}\right) \left(\exp\left(\frac{\Theta_E}{T}\right) - 1\right)^{-2}$$

$$C_{Schottky} = \frac{B}{T^2}$$

where γ is the Sommerfeld coefficient, β is a phonon specific heat parameter, $\Theta_E$ is a characteristic temperature of the low-energy Einstein mode, n is the number of atoms per formula unit, R is the gas constant, and A, B are fitting parameters. The Schottky term is described using a simplified, high-temperature limit formula.

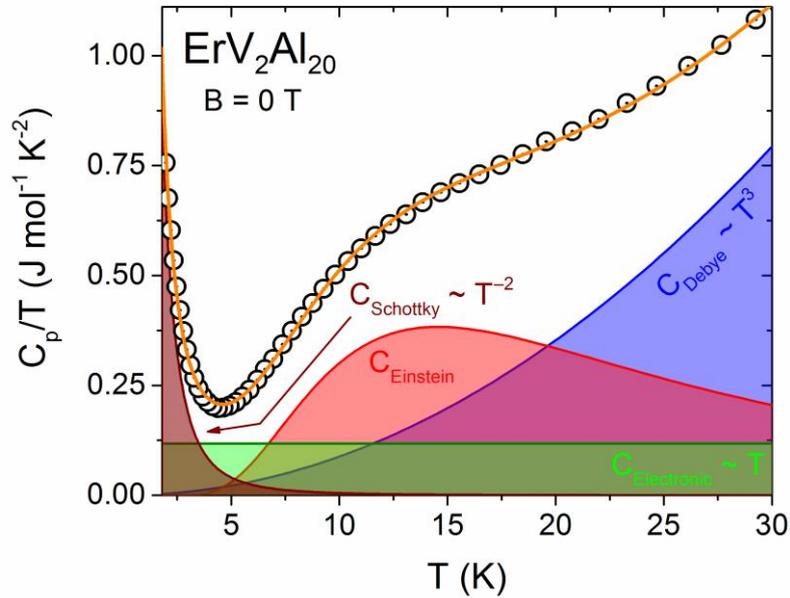

**Figure 7** Measured specific heat of ErV$_2$Al$_{20}$ at zero magnetic field (black circles). Orange line shows a fit to the experimental data including electronic, lattice, and Schottky components (see text). Contributions of individual components are plotted below: green – electronic, blue – Debye phonon heat capacity, red – low-energy Einstein phonon mode, brown – Schottky term.

The fit yields γ = 118(3) mJ mol$^{-1}$ K$^{-2}$, β = 0.882(9) mJ mol$^{-1}$ K$^{-4}$, A = 0.00294(4), $\Theta_E$ = 43.7(3) K, B = 5.1(1) J K mol$^{-1}$. The individual contributions are plotted on Figure 7. The large value of the γ coefficient compared to isostructural $R$V$_2$Al$_{20}$ ($R$ = Sc, Y, La, Lu) compounds, for which it varies from 20 to 30 mJ mol$^{-1}$ K$^{-2}$ [17], may result from an insufficient modelling of the Schottky heat capacity contribution, however some enhancement of γ is possible due to the electron-phonon coupling. The Einstein temperature obtained from the specific heat fit ($\Theta_E$) is higher than obtained from resistivity results ($\Theta_E^*$) by ca. 20%. Such underestimation of the Einstein temperature in resistivity fits was reported previously for VAl$_{10+x}$ and Ga$_x$V$_2$Al$_{20}$ compounds [16].

In order to confirm the lattice origin of the anomaly attributed to an Einstein mode, specific heat measurements were performed in magnetic fields of 0.5-9 T (main panel of Figure 8). The Schottky anomaly is shifted towards higher temperatures with an applied field due to Zeeman splitting of the Er$^{3+}$ energy levels, while the anomaly attributed to the Einstein mode is not affected. Figure 8(b) shows a fit to the specific heat at 0.5 T yielding the same Einstein temperature as obtained from the fit in zero field $\Theta_E$ = 43.8(5) K. At higher magnetic fields the Einstein contribution to specific heat is already obscured by the large Schottky anomaly.

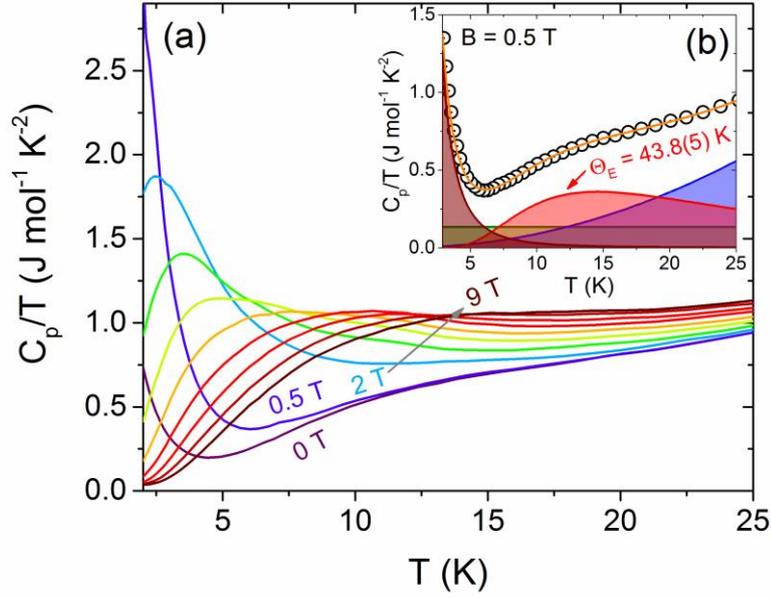

**Figure 8** (a) Specific heat of ErV$_2$Al$_{20}$ in magnetic fields B = 0.5 T and from 2 to 9 T with 1 T increments. The direction of magnetic field is parallel to the [111] crystallographic direction. Inset (b) shows a fit to the data measured at 0.5 T.

In a simple Debye model the β coefficient is related to the Debye temperature $\Theta_D$:

$$\Theta_D = \sqrt[3]{\frac{12\pi^4 nR}{5\beta}} \qquad (6)$$

The estimated value of $\Theta_D$ = 370(1) K is comparable to the value reported recently for UNb$_2$Al$_{20}$ (381 K) [55], but significantly lower than obtained for $R$V$_2$Al$_{20}$ ($R$ = Sc, Y, La, Lu) for which the Debye temperature is in the order of 500 K, and for UCr$_2$Al$_{20}$ (474 K) and ThCr$_2$Al$_{20}$ (457 K) [56]. This difference may be explained by both the effect of low-energy modes and systematic error in estimation of β parameter caused eg. by applying the low-temperature ~T$^3$ expansion of the Debye specific heat in too wide temperature range.

The fitted parameters B and γ were used to obtain a lattice specific heat (C$_{lattice}$) by subtracting the specific heat of the Schottky and electronic term from the experimental data. The results are shown in Figure 9. The peak at approx. 8.5 K confirms the presence of an Einstein mode with $\Theta_E \approx 5 \cdot 8.5 \approx 43$ K. The peak observed in ErV$_2$Al$_{20}$ is also seen in the specific heat of LaV$_2$Al$_{20}$ and LuV$_2$Al$_{20}$, however the characteristic temperature is significantly higher (ca. 140 and 90 K, respectively). A low-temperature Einstein mode was also observed in an isostructural VAl$_{10.1}$ compound with $\Theta_E$ = 21 K [13], approx. twice lower than for ErV$_2$Al$_{20}$. Since the mode in VAl$_{10.1}$ was attributed to localized anharmonic 'rattling' of Al atoms occupying the 8$a$ sites inside Al atom cages, the low-energy Einstein mode in ErV$_2$Al$_{20}$ arises likely from a 'rattling' of the cage-filling Er atoms. Relatively large thermal displacement parameters obtained from the Rietveld fit suggests that the mode could be associated with vibrations of Er(8$a$) or Al1(16$c$) atoms, both positioned inside large cages (see Figure 1(b,d)). In order to clarify this, phonon structure calculations or inelastic neutron scattering experiments should be performed.

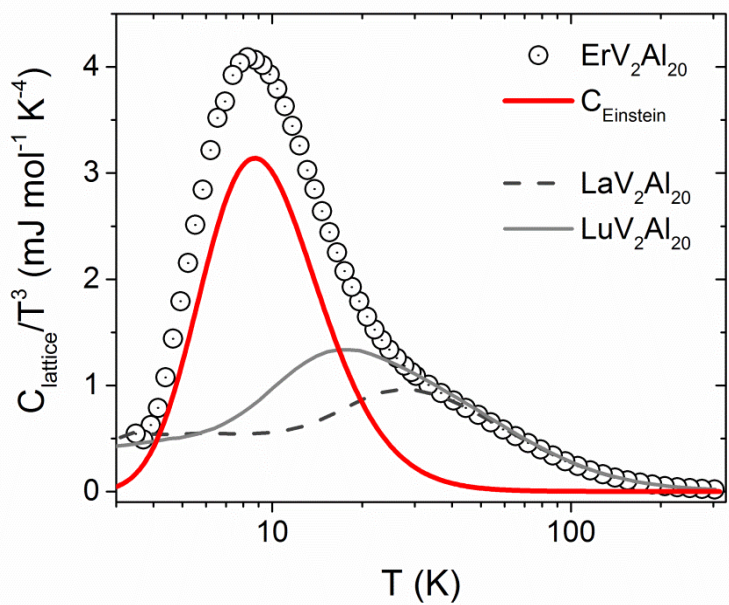

**Figure 9** Lattice specific heat of $ErV_2Al_{20}$ at zero magnetic field (black circles). Red line shows the contribution of a low-energy Einstein mode with $\Theta_E$ = 43.7 K. Lattice specific heat of $LaV_2Al_{20}$ (gray dashed line) and $LuV_2Al_{20}$ (gray solid line) are plotted for comparison after ref. [17].

**Conclusions**

Single crystals of the previously unreported $ErV_2Al_{20}$ compound crystallizing with the $CeCr_2Al_{20}$-type structure were successfully grown using the flux-growth technique. The crystal structure of the new compound was studied by means of powder X-ray diffraction and Rietveld refinement.

Magnetization measurements show Curie-Weiss paramagnetic character with an effective magnetic moment $\mu_{eff}$ = 9.27(1) $\mu_B$ close to the expected value for trivalent $Er^{+3}$ ion. The Curie-Weiss temperature $\Theta_{CW}$ = -0.55(4) K is close to zero suggesting very weak effective interactions between $Er^{3+}$ magnetic moments.

Specific heat measurements show a presence of two anomalies at low temperatures: while one, sharper, is attributed to the Schottky anomaly, the second, broader, is found to arise from the presence of a low-energy Einstein mode ($\Theta_E$ = 44 K), probably associated with the large amplitude vibrations of either Er(8*a*) or Al1(16*c*) atoms positioned in oversized cages. Therefore, $ErV_2Al_{20}$ is a new aluminide cage compound in which the 'rattling' effect is observed. Further studies, including phonon structure calculations and inelastic neutron scattering experiments, will be necessary to shed light on the origin of the low-energy Einstein mode. An interesting question to study is also how the 'rattling' of paramagnetic lanthanide atom affects the magnetic properties of the material.

**Acknowledgements**

The project was supported by the National Science Centre (Poland) grant (DEC-2012/07/E/ST3/00584).

# Crystal structure and low-energy Einstein mode in ErV$_2$Al$_{20}$ intermetallic cage compound

Michał J. Winiarski[1,*], Tomasz Klimczuk[1]

[1] *Faculty of Applied Physics and Mathematics, Gdansk University of Technology, Narutowicza 11/12, 80-233 Gdansk, Poland*

* *corresponding author: mwiniarski@mif.pg.gda.pl*

## Supplementary Material

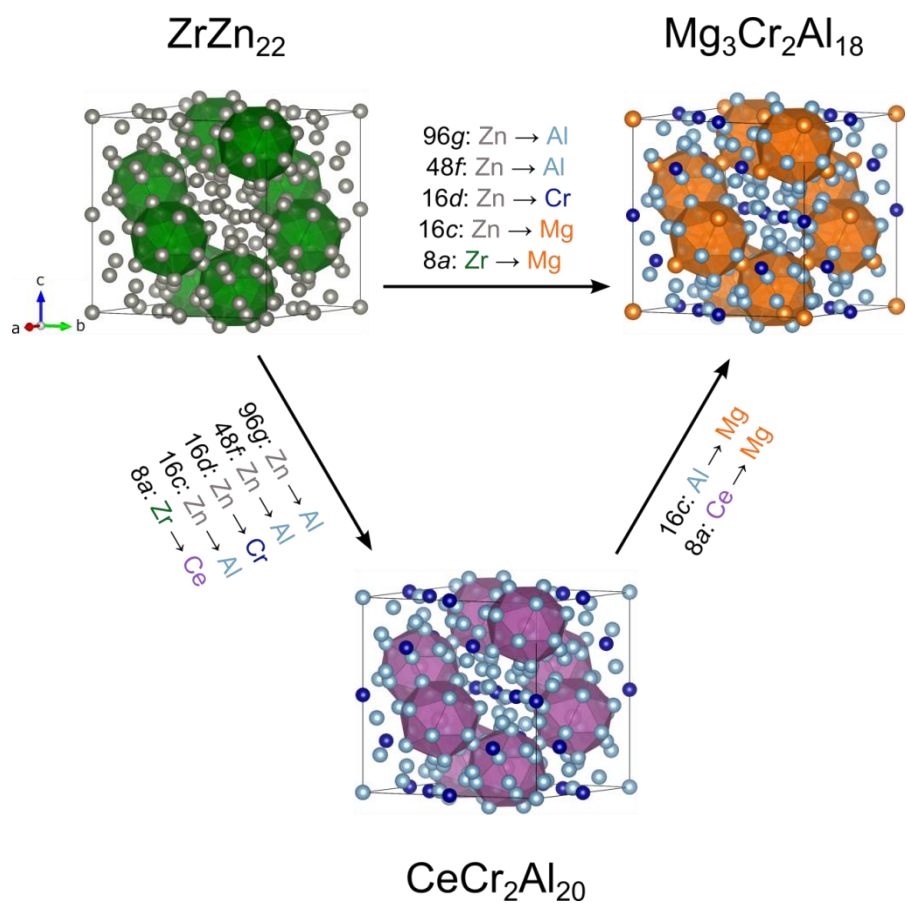

**Fig. S1: Relationships between the ZrZn$_{22}$ [1], CeCr$_2$Al$_{20}$, and Mg$_3$Cr$_2$Al$_{18}$-type [2] crystal structures.**

# Crystal structure and low-energy Einstein mode in ErV$_2$Al$_{20}$ intermetallic cage compound

Michał J. Winiarski[1,*], Tomasz Klimczuk[1]

[1] *Faculty of Applied Physics and Mathematics, Gdansk University of Technology, Narutowicza 11/12, 80-233 Gdansk, Poland*

\* corresponding author: mwiniarski@mif.pg.gda.pl

## Supplementary Material

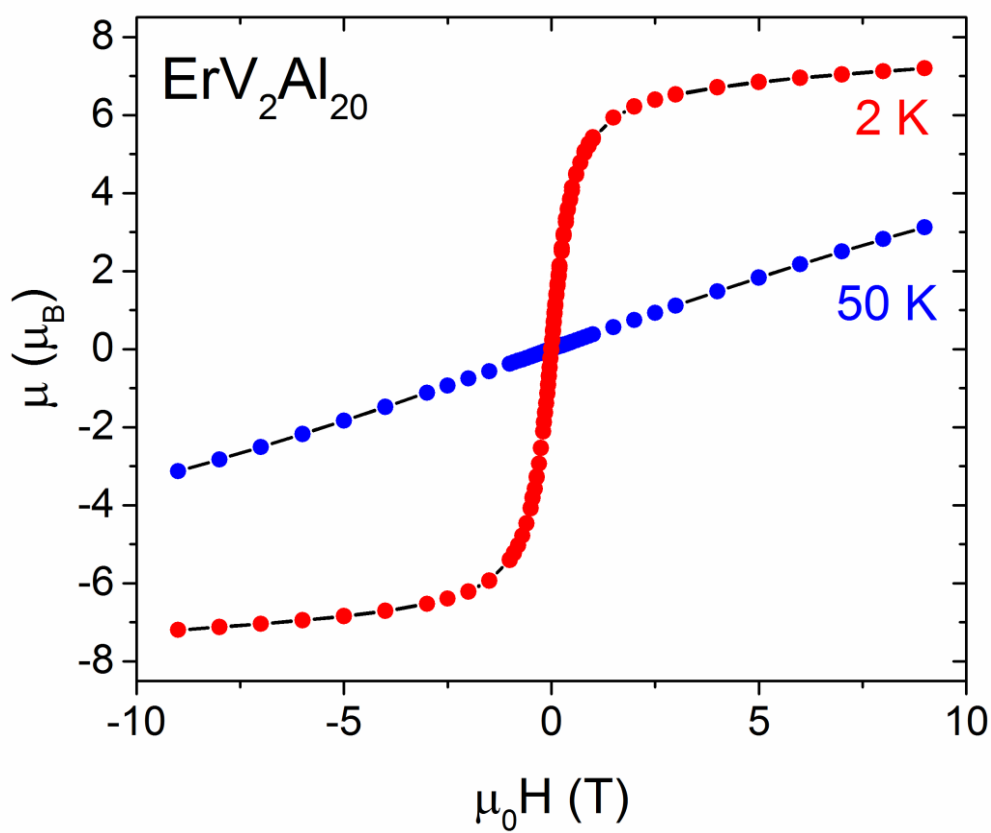

**Fig. S2:** Magnetic field dependency of magnetization for ErV$_2$Al$_{20}$ showing a linear relation at 50 K and saturating character at 2 K, as expected for a Curie-Weiss paramagnet.